\documentclass[
reprint,
showpacs,
floatfix,
aps,
prl,
nofootinbib,
twocolumn,
superscriptaddress,
amssymb,
]{revtex4-1}
\usepackage{amssymb}
\usepackage{latexsym}
\usepackage[dvips]{graphicx}
\usepackage{amsmath}
\usepackage{float}

\usepackage{graphicx}
\usepackage{dcolumn}
\usepackage{amsfonts}
\usepackage{bm}

\newcommand{\be}{\begin{equation}}
\newcommand{\ee}{\end{equation}}
\newcommand{\bea}{\begin{eqnarray}}
\newcommand{\eea}{\end{eqnarray}}

\def\bip{B_\parallel}

\def\oc{\omega_{\mbox{\scriptsize {c}}}}

\def\rxx{R_{xx}}
\def\ryy{R_{yy}}

\def\rc{R_{\mbox{\scriptsize {c}}}}

\def\easy{\left < 110 \right >}
\def\hard{\left < 1\bar{1}0 \right >}
\def\x{\hat{x}}
\def\y{\hat{y}}

\def\easy{\left < 110 \right >}
\def\hard{\left < 1\bar{1}0 \right >}

\newcommand{\rfig}[1]{Fig.\,\ref{#1}}

\newcommand{\rref}[1]{Ref.\,\onlinecite{#1}}

\begin{document}
\title{Effect of density on quantum Hall stripe orientation in tilted magnetic fields}
\author{Q.~Shi}
\affiliation{School of Physics and Astronomy, University of Minnesota, Minneapolis, Minnesota 55455, USA}
\author{M.~A.~Zudov}
\affiliation{School of Physics and Astronomy, University of Minnesota, Minneapolis, Minnesota 55455, USA}
\author{Q.~Qian}
\affiliation{Department of Physics and Astronomy and Birck Nanotechnology Center, Purdue University, West Lafayette, Indiana 47907, USA}
\author{J.\,D. Watson$^*$}
\affiliation{Department of Physics and Astronomy and Birck Nanotechnology Center, Purdue University, West Lafayette, Indiana 47907, USA}
\author{M.\,J. Manfra}
\affiliation{Department of Physics and Astronomy and Birck Nanotechnology Center, Purdue University, West Lafayette, Indiana 47907, USA}
\affiliation{Station Q Purdue, Purdue University, West Lafayette, Indiana 47907, USA}
\affiliation{School of Materials Engineering and School of Electrical and Computer Engineering, Purdue University, West Lafayette, Indiana 47907, USA}

\received{\today}

\begin{abstract}
We investigate quantum Hall stripes under in-plane magnetic field $\bip$ in a variable-density two-dimensional electron gas.
At filling factor $\nu = 9/2$, we observe one, two, and zero $\bip$-induced reorientations at low, intermediate, and high densities, respectively.
The appearance of these distinct regimes is due to a strong density dependence of the $\bip$-induced orienting mechanism which triggers the second reorientation, rendering stripes \emph{parallel} to $\bip$.
In contrast, the mechanism which reorients stripes perpendicular to $\bip$ showed no noticeable dependence on density.
Measurements at $\nu = 9/2$ and $11/2$ at the same, tilted magnetic field, allows us to rule out density dependence of the native symmetry-breaking field as a dominant factor.
Our findings further suggest that screening might play an important role in determining stripe orientation, providing guidance in developing theories aimed at identifying and describing native and $\bip$-induced symmetry-breaking fields. 
\end{abstract}

\maketitle
Quantum Hall stripe phases \cite{koulakov:1996,fogler:1996,lilly:1999a,du:1999,lilly:1999b,pan:1999,cooper:2001,zhu:2002,cooper:2004,sambandamurthy:2008,zhu:2009,kukushkin:2011,koduvayur:2011,liu:2013,liu:2013b,friess:2014,samkharadze:2016,shi:2016a,shi:2016c,pollanen:2015,mueed:2016} represent one class of exotic states that appear in a two-dimensional electron gas (2DEG) subjected to perpendicular magnetic fields and low temperatures.
These phases manifest charge clustering originating from a box-like interaction potential \cite{koulakov:1996,fogler:1996} owing to ring-shaped wavefuctions in higher Landau levels (LLs).
A built-in symmetry-breaking potential in the GaAs quantum well, hosting a two-dimensional electron gas (2DEG), macroscopically orients stripes along $\easy$ crystal direction, with very few exceptions \cite{zhu:2002,liu:2013b,pollanen:2015}.
Despite continuing efforts \cite{koduvayur:2011,sodemann:2013,liu:2013,pollanen:2015}, the origin of such preferred native orientation remains a mystery.
It is known, however, that due to a finite thickness of the 2DEG, an in-plane magnetic field $\bip$ modifies both the wavefunction and interactions which, in turn, can change stripe orientation \cite{jungwirth:1999,stanescu:2000}.

While early experiments \cite{lilly:1999b,pan:1999,cooper:2001} and theories \citep{jungwirth:1999,stanescu:2000} consistently showed that $\bip$ favors stripes perpendicular to it \citep{note:200}, subsequent studies revealed limitations of this ``standard picture''.
For example, in a tunable-density heterostructure insulated gate field effect transistor \cite{zhu:2002} native stripes along $\easy$ crystal direction did not reorient by $\bip$.
In other experiments, however, reorientation occurred even when $\bip$ was applied perpendicular to the native stripes \cite{zhu:2009,shi:2016c,note:12}.
Finally, it was recently reported that $\bip$ applied along native stripes can induce two successive reorientations, first perpendicular and then parallel to $\bip$ \cite{shi:2016c}.

Together, these experiments indicate that the impact of $\bip$ on stripe orientation remains poorly understood and is far more complex than suggested by a ``standard picture'' \cite{jungwirth:1999,stanescu:2000}.
In particular, all examples mentioned above revealed that $\bip$ can, in fact, favor \emph{parallel} (to $\bip$) stripe alignment. 
It was also found that the $\bip$-induced mechanism which favors such alignment is highly sensitive to both spin and orbital quantum numbers \citep{shi:2016c}. 
To shed light onto the nature of this mechanism, it is very desirable to identify a tuning parameter that would enable one to control stripe orientation under $\bip$.

In this Rapid Communication we study the effect of the carrier density $n_e$ on stripe orientation in a single-subband 2DEG under $\bip$ applied along native stripes ($|| \easy$).
At filling factor $\nu = 9/2$, we demonstrate three distinct classes of behavior.
At low $n_e$, we observe a single reorientation (at $\bip = B_1$) which renders stripes perpendicular to $\bip$, in agreement with the ``standard picture'' \cite{lilly:1999b,pan:1999,jungwirth:1999,stanescu:2000}.
At intermediate $n_e$, we also detect the second reorientation (at $\bip = B_2$) which reverts stripes back to their native direction, parallel to $\bip$.
Finally, at higher $n_e$ we find that $\bip$ cannot alter stripe orientation.
We further construct a phase diagram of the stripe orientation which reveals that $B_1$ is independent of $n_e$, whereas $B_2$ decreases rapidly with $n_e$ and eventually merges with $B_1$.
The appearance of the robust regime of stripes parallel to $\bip$ at higher $n_e$ can be attributed to a reduced screening due to increased inter-LL spacing.
At the same time, a density sweep at $\nu = 9/2$ and $\nu = 11/2$ at a fixed tilted magnetic field suggests that any density dependence of the native symmetry-breaking field is not an important factor in determining stripe orientation.
These findings can provide guidance to future theoretical proposals aimed at explaining parallel stripe alignment with respect to $\bip$ and identifying the native symmetry-breaking field. 

Our 2DEG resides in a 30-nm GaAs/AlGaAs quantum well (about 200 nm below the sample surface) that is doped in a 2 nm GaAs quantum well at a setback of 63 nm.
The \emph{in situ} gate consists of an $n^+$ GaAs layer situated 850 nm below the bottom of the quantum well \citep{watson:2015}.
Eight Ohmic contacts were fabricated at the corners and midsides of the lithographically-defined $1\times1$ mm$^2$ Van der Pauw mesa.
The electron density $n_e$ was varied continuously from 2.2 to $3.8 \times 10^{11}$ cm$^{-2}$.
The peak mobility was about $\mu \approx 1.2\times 10^7$ cm$^2$/Vs at $n_e \approx 3.3\times 10^{11}$ cm$^{-2}$.
Resistances $\rxx$ ($\x \equiv \hard$) and $\ryy$ ($\y \equiv \easy$) were measured by a standard low-frequency lock-in technique at temperature of about 0.1 K to avoid possible metastable orientations \cite{zhu:2002,cooper:2004}.
An in-plane magnetic field $\bip \equiv B_y$ was introduced by tilting the sample.

\begin{figure}[t]
\centering
\includegraphics{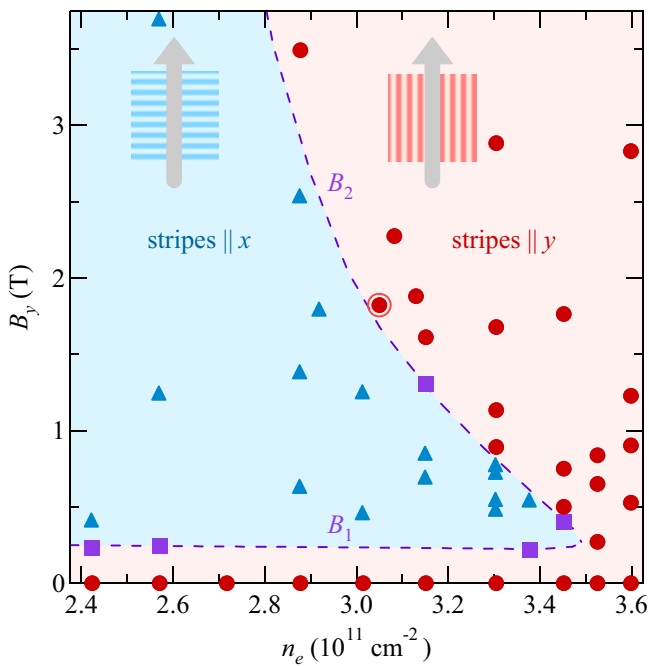}
\caption{(Color online)
Stripe orientation as a function of $n_e$ and $B_y$ at $\nu$ = 9/2. 
Triangles (circles) mark stripe orientation perpendicular (parallel) to $\bip = B_y$.
The big circle was obtained from the density sweep in \rfig{fig2}.
Squares mark the isotropic state.
The lower (upper) phase boundary (dashed line) is a guide to the eyes marking $\bip = B_1$ ($\bip = B_2$), see the text.
}
\vspace{-0.15 in}
\label{fig1}
\end{figure}

In \rfig{fig1} we summarize our experimental findings at $\nu$ = 9/2, namely the phase diagram of stripe orientation in the $(n_e,B_y)$-plane.
The diagram contains two distinct phases, ``stripes $\parallel\x$'' and ``stripes $\parallel\y$''.
While the native stripes are along the $\y$-axis at all densities studied, one easily identifies three distinct evolutions of stripe orientation with $\bip$.
At low densities we observe a \emph{single} reorientation ($\y \rightarrow \x$), in accord with the ``standard picture'' \cite{note:16,lilly:1999b,pan:1999,jungwirth:1999,stanescu:2000,cooper:2001}.
At intermediate densities, the stripes undergo \emph{two} successive reorientations ($\y \rightarrow \x$ and $\x \rightarrow \y$), ultimately aligning along $\bip$ \cite{shi:2016c}. 
Finally, the high density regime reveals \emph{no} reorientations whatsoever, and the native direction of the stripes ($\parallel y$) is preserved at all $\bip$.
While each of these regimes was previously realized in individual samples \citep{lilly:1999b,pan:1999,zhu:2002,zhu:2009,shi:2016c}, to our knowledge, it is the first observation of all three classes of behavior in a single device.

Further examination of the phase diagram (\rfig{fig1}) shows that the characteristic in-plane field $B_y = B_1$, describing the first ($\y \rightarrow \x$) reorientation, is virtually independent of $n_e$, as revealed by essentially horizontal lower boundary at $B_1 \approx 0.25$ T of the ``stripes $\parallel\x$'' phase.
On the other hand, the in-plane field $B_y = B_2$, corresponding to the second ($\x \rightarrow \y$) reorientation (the upper boundary of the ``stripes $\parallel\x$'' phase) decreases sharply with $n_e$ until it merges with $B_1$ at $n_e\approx 3.5\times 10^{11}$ cm$^{-2}$.
Indeed, $B_2$ drops by an order of magnitude over a density variation of less than 20$\,\%$.
It is this steep dependence of $B_2$ on $n_e$ that is responsible for the appearance of the three distinct regimes discussed above.

As pointed out in \rref{shi:2016c}, which investigated both $B_1$ and $B_2$ at fixed $n_e$, $B_2$ depends strongly on spin and orbital indices, in sharp contrast to $B_1$; at $\nu = 11/2$, $B_2$ is significantly higher than at $\nu = 9/2$. 
This observation, together with theoretical considerations \citep{jungwirth:1999} predicting similar $\bip$-induced anisotropy energies favoring perpendicular stripes at these filling factors, has lead to the conclusion that the second reorientation is of a different origin \citep{shi:2016c}.
The observation of strong (weak) $n_e$-dependence of $B_2$ ($B_1$) lends further support to this notion.

The $\bip$-induced anisotropy energy $E_A$ evaluated at $\bip = B_1$ is routinely used as a measure of the native anisotropy energy $E_N>0$, which aligns stripes along the $\easy$ direction at $\bip = 0$.
More specifically, the positive (negative) sign of the total anisotropy energy $E = E_N - E_A$ \citep{note:33} is reflected in the parallel (perpendicular) stripe alignment with respect to $\bip$.
Within this picture, $n_e$-independent $B_1$ suggests that $E$ is not affected by $n_e$ at $\bip \approx B_1$. 
However, $E_A$ depends on the perpendicular magnetic field $B_z$ and on the separation between subbands $\Delta$, both of which change appreciably \citep{note:201} within the density range of \rfig{fig1}.
While the exact effect of $n_e$ on $E_N$ is not known, two experiments \citep{zhu:2002,cooper:2004} revealed that $E_N$ vanishes and becomes negative above a certain $n_e$. 
In light of all these effects it is indeed surprising that $B_1$ [defined by $E_N(n_e) = E_A(B_1,n_e)$] does not depend on $n_e$ reflecting either that none of these effects is significant or that the respective changes in $E_A$ and $E_N$ compensate each other.

The rest of the phase diagram  in \rfig{fig1} clearly shows that stripe orientation is determined not by $\bip$ alone, but also by $n_e$. 
In particular, the rapid decay of $B_2$ and its merger with $B_1$ indicate that at higher $n_e$ (and higher $\bip$) stripes are more likely to be oriented parallel to $\bip$.
The decrease of $B_2$ with $n_e$, in principle, can be due to either increasing $E_N$ \citep{zhu:2002,cooper:2004} and/or decreasing $E_A$.
However, in the regime of large $\bip \gg B_1$, any change of $E_N$ is unlikely to play a big role and, as we show below, it is indeed not the driving force for the $n_e$-induced stripe reorientation observed at $\bip > B_1$ in \rfig{fig1}.

\begin{figure}[t]
\includegraphics{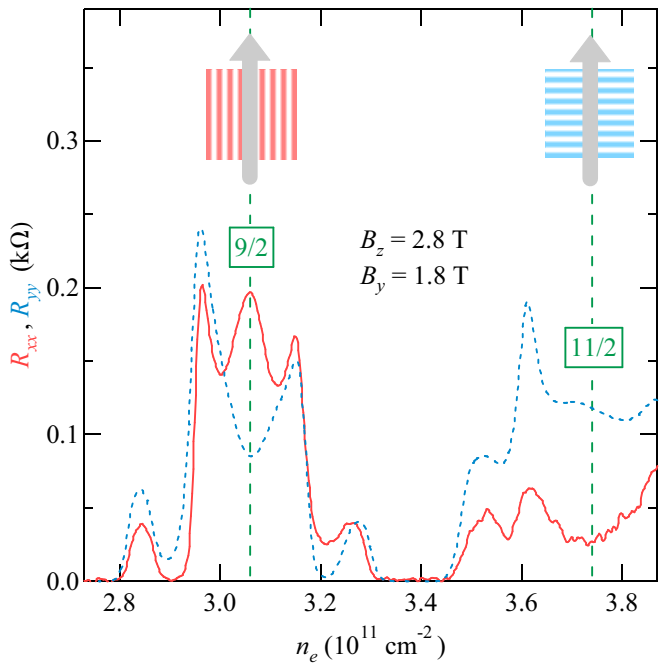}
\caption{(Color online)
$\rxx$ and $\ryy$ vs. $n_e$ measured at $B_z$ = 2.8 T and $B_y$ = 1.8 T.
}
\vspace{-0.15 in}
\label{fig2}
\end{figure}

As discussed above, $E_A$ is governed by $B_z$ and by the inter-subband splitting $\Delta$, both of which vary with $n_e$ at fixed $\nu = 9/2$, complicating the interpretation of \rfig{fig1}.
Additional information can be obtained if one fixes $B_z$ and $\bip$ and compares $\nu = 9/2$ and $11/2$ while varying $n_e$ \citep{note:999}.
To this end we have measured $\rxx$ and $\ryy$ at a fixed $B_z = 2.8$ T and $B_y = 1.8$ T while sweeping the gate voltage to cover these filling factors.
In \rfig{fig2} we present $\rxx$ (solid line) and $\ryy$ (dotted line) as a function of $n_e$.
At $\nu$ = 9/2, which occurs at a lower $n_e$, $\rxx > \ryy$ and stripes are parallel to $\y$, as a result of the second reorientation which has just occurred (cf. open circle in \rfig{fig1}).
In contrast, at $\nu = 11/2$, which is at a higher $n_e$, we find $\rxx < \ryy$ implying that stripes are still perpendicular to $\bip$.
This finding might appear puzzling as it indicates that the overall trend in \rfig{fig1}, namely that higher $n_e$ favors stripes parallel to $\bip$, is completely reversed by simply changing the spin index.

Before discussing $E_A$, we first examine if any possible density dependence of $E_N$ can explain opposite reorientation behaviors in \rfig{fig1} and \rfig{fig2}.
Increasing gate voltage (at either fixed $\nu$ or fixed $B_z$) modifies quantum confinement which can affect $E_N$, e.g. by changing the spin-orbit coupling \citep{sodemann:2013} and the strength of the interface potential experienced by electrons \citep{zhu:2002}.
However, since all the effects associated with quantum confinement are included \rfig{fig1} and \rfig{fig2} on an equal footing and they cannot be the reason for the contrasting behaviors \citep{note:22}.

Having concluded that the change of $E_N$ is of minor importance, we can now focus on $E_A$ alone.
Since the effects associated with the change of $B_z$ are absent in \rfig{fig2}, but present in \rfig{fig1}, it follows that they should play a dominant role in triggering $n_e$-induced parallel stripe alignment observed at $\bip = B_2$ in \rfig{fig1}.
One such effect is screening from other LLs which gets weaker with $B_z$ due to increased inter-LL spacing.
While theory always yields $E_A>0$ in a single-subband 2DEG when screening is taken into account, it does produce $E_A < 0$ when screening is neglected \citep{jungwirth:1999}.
One can therefore expect stripes parallel to $\bip$ if screening in realistic samples is weaker than calculations suggest \citep{aleiner:1995,kukushkin:1988,note:103}.
While there are other $B_z$-related effects that might affect $E_A$, ($\x \rightarrow \y$) stripe reorientation with increasing $n_e$ in \rfig{fig1} can be explained qualitatively by decreasing screening which favors stripes parallel to $\bip$.

On the other hand, what exactly drives the reorientation in \rfig{fig2} is not clear.
Since $E_A$ relies on the finite thickness of the 2DEG, the initial decrease of $\Delta$ enhances $E_A$ \citep{jungwirth:1999,stanescu:2000}, in agreement with recent measurements of $B_1$ \citep{pollanen:2015}.
However, when the valence LL is sufficiently close to the second subband (i.e., when $\hbar\oc/\Delta$ is slightly below 0.5 at $\nu = 9/2$ or 11/2), the system becomes more akin to a two-subband system resulting in a lower $E_A$ \citep{cooper:2001,jungwirth:1999,note:01}.
The $n_e$-driven ($\y \rightarrow \x$) reorientation of stripes in \rfig{fig2} implies that $E$ increases with decreasing $\Delta$.
However, judging what happens to $E_A$ based on theoretical calculations \citep{jungwirth:1999} is not possible because a decrease of $E_A$ with $\bip$, observed at both $\nu =9/2$ and 11/2 under the conditions of \rfig{fig2}, is not anticipated in a single-subband system \citep{jungwirth:1999}.
In addition, as discussed above, $E_N$ might also change in the density sweep.

It remains to be understood why stripes parallel to $\bip$ in a single-subband quantum well are never predicted by theories \cite{jungwirth:1999} that calculate the dielectric function using the random phase approximation (RPA) \citep{aleiner:1995}.
While we cannot point out the exact reason, it appears plausible that such calculations might not accurately capture a real experimental situation.
For example, the period of the stripe phase might be different from what Hartree-Fock calculations suggest. 
Experiments employing surface acoustic waves have obtained about a $30\%$ larger stripe period than suggested by theory \cite{koulakov:1996,fogler:1996}.
In addition, LL mixing effects beyond the RPA or disorder-induced LL broadening \cite{sodemann:2013} were not taken into account.

While the phase diagram of stripe orientation shown in \rfig{fig1} clearly identifies a robust regime of stripes parallel to $\bip$, it would indeed be interesting to extend studies to even higher carrier densities without populating the second subband.
In particular, it might allow observation of all three distinct regimes at other filling factors, e.g. $\nu = 11/2, 13/2, 15/2$.
In addition, higher $n_e$ might reveal a regime of native stripe orientation along the $\hard$ crystal direction which might allow us to establish a connection, if any, with the findings of \rref{zhu:2002}.

In summary, we have studied the effect of the carrier density $n_e$ on stripe orientation in a single-subband 30 nm-wide GaAs quantum well under $\bip$ applied along native stripes ($\parallel \easy$).
At filling factor $\nu = 9/2$, we have observed one, two, and zero $\bip$-induced stripe reorientations at low, intermediate, and high density, respectively.
The in-plane magnetic field $\bip = B_1$, which reorients stripes perpendicular to it in accord with the ``standard picture'' \cite{lilly:1999b,pan:1999,jungwirth:1999,stanescu:2000}, changes only slightly, if at all, over a wide range of densities.
In contrast, the second characteristic field $\bip = B_2$, which renders stripes parallel to $\bip$, rapidly decays with density eventually merging with $B_1$.
The observation that increasing carrier density promotes stripes parallel to $\bip$ can be qualitatively ascribed to a weaker screening due to increased inter-LL spacing which can reduce $\bip$-induced anisotropy energy and even change its sign \citep{jungwirth:1999}.
At the same time, our data suggest that the density dependence of the native symmetry-breaking field, if any, is not an important factor in determining stripe orientation.
Our findings provide guidance to future theories attempting to explain parallel stripe alignment with respect to $\bip$ and to identify the native symmetry-breaking field. 
These theories should also take into account experimental evidence \citep{zhu:2002,zhu:2009,shi:2016c} for anisotropic nature of $E_A$.

\begin{acknowledgements}
We thank I. Dmitriev, A. Kamenev, B. Shklovskii, and I. Sodemann for discussions and H. Baek, G. Jones, S. Hannas, T. Murphy, and J. Park for technical assistance.
The work at Minnesota (Purdue) was supported by the U.S. Department of Energy, Office of Science, Basic Energy Sciences, under Award \# ER 46640-SC0002567 (DE-SC0006671).
A portion of this work was performed at the National High Magnetic Field Laboratory, which is supported by National Science Foundation Cooperative Agreement No. DMR-1157490 and the State of Florida.
\end{acknowledgements}
\small{$^*$Current address: QuTech and Kavli Institute of Nano\,-science, Delft Technical University, 2600 GA Delft, The Netherlands.}


\begin{thebibliography}{39}
\bibitem[{not({\natexlab{a}})}]{note:200}
\bibinfo{note}{We limit our discussion to single-subband 2DEG.}
\bibitem[{not({\natexlab{d}})}]{note:16}
\bibinfo{note}{We limit the range of $B_y$ to avoid complications from
  LL-mixing effects.}
\bibitem[{not({\natexlab{g}})}]{note:999}
\bibinfo{note}{$\nu = 9/2$ and $11/2$ are characterized by the same 2D
  wavefunction, stripe period, and inter-LL spacing, justifying their direct
  comparison in \rfig{fig2}.}
\bibitem[{not({\natexlab{h}})}]{note:22}
\bibinfo{note}{At a fixed $\nu$, larger $n_e$ also translates to a reduced
  stripe period which can also affect $E_N$. However, \rref{zhu:2002} found
  that $E_N$ vanishes at both $\nu = 9/2$ and 11/2 at the same $n_e$,
  indicating that this effect is of minor importance.}
\bibitem[{not({\natexlab{i}})}]{note:103}
\bibinfo{note}{Screening at a wavevector of stripe modulation is characterized
  by $\varepsilon (1+a/\pi a_B)$, where $\varepsilon \approx 12.8$ is the
  dielectric constant of GaAs, $a \approx 2.8 \rc$ is the stripe period, $\rc$
  is the cyclotron radius, and $a_B \approx 10$ nm is the Bohr radius
  \citep{aleiner:1995,koulakov:1996,kukushkin:1988}.}
\end{thebibliography}
\end{document}